\DocumentMetadata{}
\documentclass[sigconf,noacm]{acmart}

\AtBeginDocument{%
  \providecommand\BibTeX{{%
    \normalfont B\kern-0.5em{\scshape i\kern-0.25em b}\kern-0.8em\TeX}}}

\makeatletter
\def\@ACM@checkaffil{
    \if@ACM@instpresent\else
    \ClassWarningNoLine{\@classname}{No institution present for an affiliation}%
    \fi
    \if@ACM@citypresent\else
    \ClassWarningNoLine{\@classname}{No city present for an affiliation}%
    \fi
    \if@ACM@countrypresent\else
        \ClassWarningNoLine{\@classname}{No country present for an affiliation}%
    \fi
}
\makeatother

\acmYear{2022}

\acmConference[UW--Madison]{UW--Madison}{January 05, 2023}{Madison, WI}
\setcopyright{none}
\renewcommand\footnotetextcopyrightpermission[1]{} 

\usepackage{tikz}
\usetikzlibrary{graphs,quotes}

\usepackage{amsmath}
\usepackage{centernot}
\usepackage{xspace}
\usepackage{xcolor}
\usepackage{multirow}
\usepackage{wasysym}

\usepackage{caption}
\usepackage{enumitem}
\setlist[itemize]{leftmargin=10pt}

\newcommand{\paratitle}[1]{\vspace{5pt}\noindent\textbf{#1}.}

\newcommand{\opr}[3]{$|#1\text{R}#2{:}#3|$}
\newcommand{\opw}[3]{$|#1\text{W}#2{\angle}#3|$}
\newcommand{\oprmw}[4]{$|#1\text{RMW}#2{:}#3{\angle}#4|$}

\newcommand{\ptopr}[2]{$#1\text{R}#2$}
\newcommand{\ptopw}[3]{$#1\text{W}#2{\angle}#3$}

\newcommand{\nleadsto}[0]{\centernot\leadsto}
\newcommand{\unordered}[0]{\centernot\leftrightsquigarrow}

\newcommand*\circleb[1]{\tikz[baseline=(char.base)]{
    \node[shape=circle,fill,inner sep=.5pt](char){\textcolor{white}{\small #1}};}}

\makeatletter
\newcommand{\thefontsize}{\f@size pt}
\makeatother

\begin{document}

\title{A Unified, Practical, and Understandable Model of Non-transactional Consistency Levels in Distributed Replication}

\author{Guanzhou Hu}
\affiliation{\institution{UW--Madison}}
\email{guanzhou.hu@wisc.edu}

\author{Andrea C. Arpaci-Dusseau}
\affiliation{\institution{UW--Madison}}
\email{dusseau@cs.wisc.edu}

\author{Remzi H. Arpaci-Dusseau}
\affiliation{\institution{UW--Madison}}
\email{remzi@cs.wisc.edu}


\begin{abstract}
We present a practical model of non-transactional \textit{consistency} levels in the context of distributed data replication. Unlike prior work, our simple Shared Object Pool (SOP) model defines common consistency levels in a unified framework centered around the single concept of \textit{ordering}. This naturally reflects modern cloud object storage services and is thus easy to understand. We show that a consistency level can be intuitively defined by specifying two types of constraints on the validity of orderings allowed by the level: \textit{convergence}, which bounds the lineage shape of the ordering, and \textit{relationship}, which bounds the relative positions between operations. We give examples of representative protocols and systems, and discuss their \textit{availability} upper bound. To further demonstrate the expressiveness and practical relevance of our model, we use it to implement a Jepsen-integrated consistency checker for the four most common levels (linearizable, sequential, causal+, and eventual); the checker analyzes consistency conformity for small-scale histories of real system runs (etcd, ZooKeeper, and RabbitMQ).
\end{abstract}

\settopmatter{printfolios=true}
\maketitle
\pagestyle{plain}

\section{Introduction}

A crucial step towards designing distributed replication protocols and building reliable distributed storage systems is to define their consistency semantics\footnote{By \textit{consistency}, we refer to the constraints that restrict which orderings of operations on shared data objects are considered valid, as defined in \S\ref{sec:problem-model}. This is not to be confused with the ``C'' property in transactional ACID properties~\cite{transaction-concept, transaction-oriented-recovery}, which refers to application-level integrity invariants. In fact, consistency in our context maps to the ``I'' (\textit{isolation}) property in ACID, as we explain in \S\ref{sec:problem-model}.}. However, apart from the purely formal summary by Viotti and Vukolić~\cite{non-transactional-summary}, there has been no unified definition of existing consistency levels in the context of distributed replication systems. This is largely due to the rich history of research that contributed to this field. Many fundamental breakthroughs stemmed from different research areas, including distributed system modeling~\cite{logical-clocks, linearizability, sequential-consistency, consistency-availability-convergence, session-guarantees, paxos-parliament, synthesis-lectures}, multiprocessor shared-memory consistency~\cite{multiprocessors-simple-consistency-models, memory-consistency-models, shared-memory-consistency, parallel-memory-consistency-survey, causal-memory, processor-consistency, primer-memory-consistency}, network reliability modeling~\cite{cap-theorem, towards-robust-distributed, flp-impossibility, why-do-computers-stop, pbft}, and database transaction processing~\cite{transaction-concept, transaction-oriented-recovery, transaction-in-r-star}. They discuss different pieces of the problem within different contexts, leading to plentiful but sometimes blurry terminology when applied to distributed replication.

We propose a minimal yet self-contained theoretical framework -- the Shared Object Pool (SOP) model -- which unifies the definition of common consistency levels in a way that is understandable to protocol designers and system engineers. We restrict our discussion to a selected set of non-transactional consistency levels seen in real object storage designs. To further improve understandability, we use examples extensively to explain the practical differences between consistency levels, and refer to representative protocols and systems corresponding to each level.

The rest of the paper is organized as follows. \S\ref{sec:problem-model} describes our problem model setup, defines ordering, and explains the meaning of non-transactional consistency within this context. \S\ref{sec:ordering} defines all variants of ordering validity constraints. \S\ref{sec:consistency-levels} presents the hierarchy of selected consistency levels, dissects their ordering validity guarantees, explains their practical differences, and gives examples of representative protocols and systems. \S\ref{sec:availability-guarantees} discusses the availability upper bounds in the presence of network partitioning. \S\ref{sec:checker-impl} presents a checker prototype implementation. \S\ref{sec:conclusion} concludes.

\section{Problem Model}
\label{sec:problem-model}

We model our problem setup as a conceptual object storage service, which we term a Shared Object Pool (SOP). In this section, we define the SOP model and explain the meaning of consistency.

\begin{figure}[t]
    \centering
    \vspace*{18pt}
    \includegraphics[width=0.88\columnwidth]{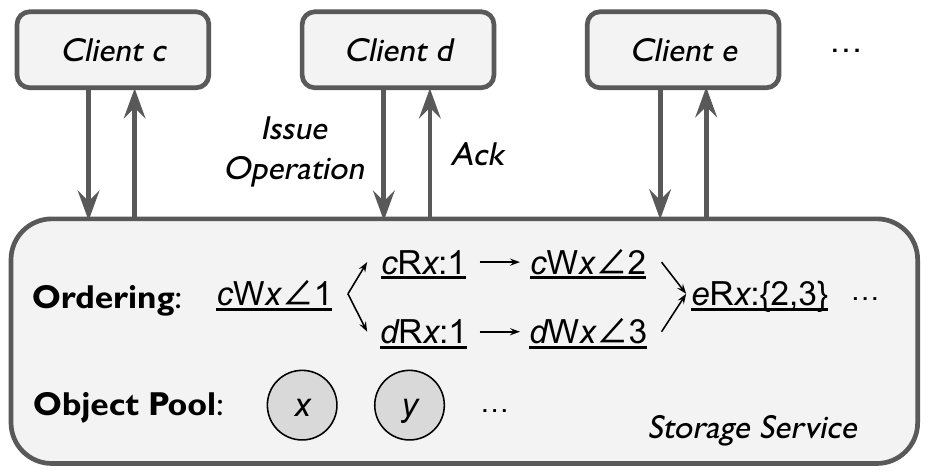}

    \vspace{-5pt}
    \caption{Shared Object Pool (SOP) Model.}
    \vspace{-12pt}
    \label{fig:sop-model}
\end{figure}

\subsection{Shared Object Pool (SOP) Model}
\label{sec:sop-model}

We consider a storage \textit{service} shared by multiple \textit{clients}, as shown in Figure~\ref{fig:sop-model}. The service appears to be a pool of \textit{objects}. Each object has a unique name and contains a \textit{value}; it is a \textit{register} in classic literature. The only way to learn about an object's value is through the result of a read operation, which we introduce below. Objects are not necessarily stored as physical bytes on physical machines; the SOP model is entirely conceptual and is agnostic to any actual design of protocols and implementation of systems.

Clients are single-threaded, closed-loop entities that invoke \textbf{operations} on the service. When a client $c$ issues an operation $p$, it blocks until the acknowledgment of $p$ by the service. Multi-threaded or asynchronous client implementations should be modeled as multiple SOP clients. An operation is of one of the following types:
\begin{itemize}
    \item \textbf{Read} (R): we use \opr{c}{x}{v} to denote client $c$ reading object $x$ and getting the result value $v$ upon acknowledgement. A read operation may also return a set of values, or some arbitrarily reduced value by applying a function $f$ to a set of values. We denote this as \opr{c}{x}{f(\{v_1, v_2\})}, or just \opr{c}{x}{\{v_1, v_2\}} for short.
    \item \textbf{Write} (W): we use \opw{c}{x}{v} to denote client $c$ overwriting object $x$'s value with value $v$.
    \item \textbf{Read-Modify-Write} (RMW): we use \oprmw{c}{x}{v}{v'} to denote a compound read-modify-write operation on object $x$, which reads the value of $x$, getting $v$, and writes back a new value $v'$ based on some arbitrary computation over the result of the read. One representative RMW operation is \textit{conditional write}, e.g., \textit{compare-and-swap} (CAS), which reads the current value, compares it against a given value $v$, and writes a new value $v'$ if the comparison shows equality or writes $v' = v$ back otherwise.
\end{itemize}

The types of objects and operations can be generalized. For example, objects can be counters or queues, and RMW operations can be extended to arbitrary commands. We use the above read-write-style definition throughout this paper for clarity.

The service maintains a possibly partial \textbf{ordering} $O$ of all operations acknowledged. The ordering $O$ captures dependencies between operations enforced by the service and materializes the result of each operation. Given a \textit{workload} of operations generated by clients, whether an ordering is acceptable or not is decided by some \textit{validity constraints}. Modeling the validity constraints guaranteed by the service effectively models its interface semantics, hence its \textit{consistency level}. The following three subsections explain the meaning of workload, ordering, and consistency, respectively.

\subsection{Physical Timeline Workload}
\label{sec:physical-timeline}

In the SOP model, each client is a single-threaded entity. For a concrete collection of client operations, we can visualize the \textit{physical timeline} $T$ of when each operation is issued and acknowledged. Every row represents a client, while the x-axis represents the real-world time at which an operation is issued or acknowledged.

For example, below is a physical timeline of two clients, $c$ and $d$, performing operations on two objects, $x$ and $y$:
\begin{center}
\tikz \graph [no placement, empty nodes, nodes={circle, fill, scale=0.3}] {
    c1[x=-0.5,y=0,fill=white,scale=3.5,as=$c:$];
    a[x=0,y=0];
    b[x=1.5,y=0];
    c[x=2.5,y=0];
    d[x=3.5,y=0];
    e[x=4.8,y=0];
    f[x=5.8,y=0];

    c2[x=-0.5,y=-0.8,fill=white,scale=3.5,as=$d:$];
    g[x=1,y=-0.8];
    h[x=2.2,y=-0.8];
    i[x=3.2,y=-0.8];
    j[x=4.5,y=-0.8];
    
    a --["\ptopw{c}{x}{1}"] b;
    c --["\ptopw{c}{x}{3}"] d;
    e --["\ptopr{c}{y}"] f;

    g --["\ptopr{d}{x}"] h;
    i --["\ptopw{d}{y}{2}"] j;
};
\end{center}

A physical timeline depicts a concrete history of client activity. We can think of it as a specific ``workload'' that drives the storage service. Given a physical timeline, the storage service delivers a final ordering (from the set of valid orderings allowed by its consistency level) that connects all operations in the timeline together.

Results of read values in R and RMW operations are \textit{not} part of the physical timeline workload. Rather, they are materialized in the final ordering decided by the service.  Everything else about client operations activity is included in the physical timeline.

Values of writes are part of the workload. Although we use concrete numeric values as examples throughout this paper, they can also be symbolic values that capture the program logic of client applications. For instance, \opw{d}{y}{2} in the example above may instead be \opw{d}{y}{v}, where $v$ is a symbolic value that represents applying some function over the return value of $d$'s preceding read of object $x$. The write value of an RMW operation is typically a symbolic value that depends on the result of the read.

\subsection{Definition of Ordering}

An \textit{ordering} is a \textit{directed acyclic graph} (DAG), where nodes are operations from a physical timeline workload. Each operation that has been acknowledged appears exactly once in an ordering. Pending operations that have not been acknowledged are not interesting in our definition of consistency and are thus not explicitly discussed. A directed edge connecting two operations represents an ``ordered before'' relationship between the two.

We say an operation $op_1$ is \textit{ordered before} $op_2$ (denoted $op_1 \leadsto op_2$) in ordering $O$ iff. there exists either an edge in $O$ pointing from $op_1$ to $op_2$, or an operation $op'$ such that $op_1 \leadsto op'$ and $op' \leadsto op_2$ (transitivity). If neither operation is ordered before the other, that is, $op_1 \nleadsto op_2$ and $op_2 \nleadsto op_1$, then we say $op_1$ and $op_2$ are \textit{unordered} with each other (denoted $op_1 \unordered op_2$).

Given a physical timeline, an ordering is \textit{valid} on the timeline with respect to a consistency level if it satisfies the validity constraints enforced by that level.

\paratitle{Early Literature Terminology} Similar definitions of ``ordered before'' relationship have appeared in many early literature~\cite{logical-clocks, sequential-consistency, linearizability, sequential-vs-linearizability, replicated-data-types}, where it was termed ``happens before'' and was associated with single-point \textit{events}. Unordered events in a partial ordering were often termed ``concurrent'' events. In this paper, we use ``ordered before'' and ``happens before'' interchangeably, and use ``unordered'' and ``concurrent'' interchangeably, but on operations.

\subsection{Meaning of Consistency}
\label{sec:meaning-of-consistency}

The \textbf{consistency level} of the storage service is determined by \textit{which orderings of operations are considered valid} given any physical timeline workload. In other words, the consistency level enforces \textit{what validity constraints must be held} on the ordering given any workload. A stronger consistency level imposes more constraints than a weaker one and therefore disallows more orderings, exposing an interface that is more restrictive in the protocol design space and in the meantime easier to use by clients. In contrast, a weaker consistency level relaxes certain constraints and opens up new opportunities in the protocol design space, albeit providing weaker semantic guarantees for clients.

An ordering represents \textit{logical} dependencies among operations, similar to Lamport's definition of logical clock on events~\cite{logical-clocks}, and does not necessarily capture physical time; in fact, whether physical time is respected or not is one of the validity constraints that differentiate several consistency levels.
Our SOP model shares similarities with the specification framework for replicated data types proposed by Burckhardt et al.~\cite{replicated-data-types}; the differences are that we simplify the notion of ordering and cover stronger consistency levels (rather than focusing only on causal and eventual consistency models).

Note that the SOP model is oblivious to any system design and implementation details of the service, including but not limited to how the service is constructed out of servers, what the network topology looks like, and how client-server connections are established. These internal design choices should not affect the interface semantics exposed to clients.

We only consider a \textit{non-transactional} storage service interface, where each operation touches exactly \textit{one} object. Transactional operations, which group multiple single-object operations together, open up a new dimension in the consistency level space and are essential to distributed database systems. A common practice in modern database systems is to deploy \textit{sharded concurrency control} mechanisms atop replicated data objects, effectively layering transactional guarantees separately from single-object consistency~\cite{cockroachdb, tidb, scylladb}. Despite this, transaction isolation levels can indeed be integrated into the same unified theoretical framework with single-object consistency as seen in previous literature~\cite{highly-available-transactions, jepsen-website} (because they are both rooted in the validity of orderings). We leave such integration into the SOP model as future work.

\paratitle{Early Literature Terminology} In early literature on shared memory consistency, operations are further decomposed into \textit{events}~\cite{linearizability}. The invocation and acknowledgment of an operation are considered two separate events. All events form a strictly serial sequence, named a \textit{history}. Consistency levels are then defined on the validity of well-formed histories. In this paper, we simplify this notation and choose not to use the words ``event'' and ``history''. Instead, we take a different approach and consider each operation $op$ as a contiguous timespan from its start (when the client issues $op$) to its end (when the service acknowledges $op$ and returns a result to the client). When discussing ordering of operations, we use partial ordering to depict incomparability if necessary, instead of merging them into a serial history of events. We found this approach easier to understand and visualize.

\section{Ordering Validity Constraints}
\label{sec:ordering}

In this section, we list two sets of \textit{validity constraints} that determine which orderings are acceptable in a consistency level. Specifically, the two sets are: 1) \textit{convergence constraints}, which bound the lineage ``shape'' of the ordering, and 2) \textit{relationship constraints}, which bound the ``placement'' of operations with respect to each other within the ordering given any physical timeline workload.

\subsection{Convergence Constraints}

The convergence constraints restrict whether a valid ordering must be a serial order or can be a partial order, and in the latter case, whether reads must observe convergent results. The three levels of convergence constraints are, from the strongest to the weakest, \textit{Serial Order} (SO), \textit{Convergent Partial Order} (CPO), and \textit{Non-convergent Partial Order} (NPO).

\subsubsection{Serial Order (SO)}

An SO ordering must be a \textit{total order} of operations, forming a single serial chain. 

The result of a read (or RMW) on object $x$ is determined by the latest write (of RMW) operation that \textit{immediately precedes} the read. We say an operation $op_1$ immediately precedes operation $op_2$ iff.:
\begin{itemize}
    \item they are on the same object $x$, and
    \item $op_1 \leadsto op_2$, and
    \item there is no other write (or RMW) operation $op'$ on object $x$ s.t. $op_1 \leadsto op' \leadsto op_2$.
\end{itemize}

\noindent If there is no immediately-preceding operation for a read, we assume a special initial value, e.g. 0, for every object.

Below is an example ordering that satisfies SO:
\begin{center}
\tikz \graph [grow right sep=0.5cm] {
    a/\opw{c}{x}{1} -> b/\opw{d}{x}{2} -> c/\opr{c}{x}{2} -> d/\opw{d}{y}{2} -> e/\opr{c}{y}{2};
};
\end{center}

SO is the strongest convergence constraint that any consistency level can enforce. Every operation has a relative position w.r.t. any other operation in the total order (with the exception of a cluster of pure reads shown below). It implies that the service must maintain a centralized view, e.g. a \textit{log}, of all operations~\cite{paxos-parliament, paxos-made-simple}; an operation from a client can never be acknowledged solely on its own will.

\paratitle{Cluster of Reads} We make one exception to the seriality of operations in an SO ordering: any cluster of pure read operations in between two writes are allowed to be unordered with each other. For example, the following ordering is a valid SO ordering:

\begin{center}
\tikz \graph [grow right sep=0.5cm] {
    a/\opw{c}{x}{1} -> b/\opw{c}{y}{2} -> {c/\opr{c}{x}{1}, d/\opr{d}{y}{2}, e/\opr{e}{x}{1}} -> f/\opw{c}{x}{3};
};
\end{center}

\noindent Without loss of generality, in this paper, we always present a serial chain when giving SO ordering examples for clarity.

\subsubsection{Convergent Partial Order (CPO)}

A CPO ordering can be a \textit{partial order} of operations. Writes may be unordered with some other operations, forming branches.

In addition, the result of a read must be \textit{strongly convergent}~\cite{non-transactional-summary}, meaning that it must observe all operations to the same object that immediately precede it. If multiple operations with different values to the same object all immediately precede the read and they are unordered with each other, then the read must return the set of all these values (or a reduced value over the set by applying a \textit{convergent} reduction function, as described in \S\ref{sec:sop-model}).

Below is an example ordering that satisfies CPO (but not SO):
\begin{center}
\tikz \graph [grow right sep=0.45cm] {
    a/\opw{c}{x}{1} -> b/\opr{d}{x}{1} -> { c/\opw{c}{y}{2} -> d/\opw{c}{y}{3}, e/\opw{d}{y}{4} } -> {f/\opr{e}{y}{\{3,4\}}, g/\opr{f}{y}{\{3,4\}}};
    d -> g;
    e ->[bend right=8] f;
};
\end{center}

\noindent Notice how certain operations are unordered with each other, e.g., \opw{c}{y}{2} $\unordered$ \opw{d}{y}{4} and \opw{c}{y}{3} $\unordered$ \opw{d}{y}{4}. Also notice that \opr{e}{y}{\{3,4\}} and \opr{f}{y}{\{3,4\}} must observe both values 3 and 4.

CPO opens the opportunity to allow temporarily diverging states of object values, as long as they collapse into a convergent state at some read. This typically gives protocol designers more space to improve the scalability and availability of the service.

\subsubsection{Non-convergent Partial Order (NPO)}

An NPO ordering can be a partial order of operations, just like in CPO. Furthermore, reads (and RMWs) do not have to be convergent. They are allowed to only observe a \textit{subset} of values from immediately-preceding operations, or apply a \textit{diverging} reduction function that may produce different values on different clients given the same set of input values. Reads still have to be \textit{well-formed}, meaning they cannot observe values that come from nowhere\footnote{For more complex object types such as counters or queues, this means values observed must all obey \textit{return value consistency} of the object semantic~\cite{non-transactional-summary}. We assume return value consistency is held for all consistency levels discussed in this paper, as is the case in all practical cloud systems.}.

Below is an example ordering that satisfies NPO (but not CPO):
\begin{center}
\tikz \graph [grow right sep=0.45cm] {
    a/\opw{c}{x}{1} -> b/\opr{d}{x}{1} -> { c/\opw{c}{y}{2} -> d/\opw{c}{y}{3}, e/\opw{d}{y}{4} } -> {f/\opr{e}{y}{3}, g/\opr{f}{y}{4}};
    d -> g;
    e ->[bend right=8] f;
};
\end{center}

\noindent Notice that \opr{e}{y}{3} is now allowed to only observe value 3 and miss the existence of value 4; similarly for \opr{f}{y}{4}.

NPO allows clients to observe forever-diverging values of the same object. Without careful assistance from the relationship constraints side, a service that only guarantees NPO can hardly provide any reasonable consistency semantic.

\subsection{Relationship Constraints}

The relationship constraints restrict how operations are placed with respect to each other in the final ordering. More specifically, they determine what properties of the physical timeline workload must be reflected in the ordering. The four levels of relationship constraints are, from the strongest to the weakest, \textit{Real-Time} (RT), \textit{Causal} (CASL), \textit{First-In-First-Out} (FIFO), and \textit{None}.

\subsubsection{Real-Time (RT)}
\label{sec:relationship-rt}

In an RT ordering, if operation $op_1$ ends before operation $op_2$ starts in \textit{physical time} (regardless of whether they come from different clients or are on different objects), then the ordering must enforce $op_1 \leadsto op_2$.

For example, given the physical timeline below:
\begin{center}
\tikz \graph [no placement, empty nodes, nodes={circle, fill, scale=0.3}] {
    c1[x=-0.5,y=0,fill=white,scale=3.5,as=$c:$];
    a[x=0,y=0];
    b[x=1.2,y=0];
    c[x=1.5,y=0];
    d[x=2.7,y=0];

    c2[x=-0.5,y=-0.8,fill=white,scale=3.5,as=$d:$];
    i[x=3,y=-0.8];
    j[x=3.8,y=-0.8];
    k[x=4.2,y=-0.8];
    l[x=5.4,y=-0.8];

    c3[x=-0.5,y=-1.6,fill=white,scale=3.5,as=$e:$];
    g[x=0.7,y=-1.6];
    h[x=1.9,y=-1.6];
    m[x=5.7,y=-1.6];
    n[x=6.5,y=-1.6];
    
    a --["\ptopw{c}{x}{1}"] b;
    c --["\ptopw{c}{x}{2}"] d;

    i --["\ptopr{d}{x}"] j;
    k --["\ptopw{d}{y}{3}"] l;
    
    g --["\ptopw{e}{x}{3}"] h;
    m --["\ptopr{e}{y}"] n;
};
\end{center}

The following is an ordering that is SO and RT:
\begin{center}
\tikz \graph [grow right sep=0.2cm] {
    a/\opw{c}{x}{1} -> b/\opw{e}{x}{3} -> c/\opw{c}{x}{2} -> d/\opr{d}{x}{2} -> e/\opw{d}{y}{3} -> f/\opr{e}{y}{3};
};
\end{center}

And the following is an ordering that is CPO and RT:
\begin{center}
\tikz \graph [grow right sep=0.32cm] {
    {a/\opw{c}{x}{1} -> b/\opw{c}{x}{2}, c/\opw{e}{x}{3}} -> d/\opr{d}{x}{\{2,3\}} -> e/\opw{d}{y}{3} -> f/\opr{e}{y}{3};
};
\end{center}

RT is the strongest relationship constraint that any consistency level can enforce. For each client, its operations exhibit the same order as how the client issues them, because an operation naturally finishes before the start of the next one following it on the same client. Across different clients, RT ensures that an operation observes all other operations acknowledged before its start.

The RT guarantee implies that the service must deploy some synchronization mechanism across all clients; an operation from a client can never be acknowledged solely on its own will.

\subsubsection{Causal (CASL)}
\label{sec:relationship-casl}

The causal guarantee relaxes RT by allowing more cases of reordering between cross-client operations. If operation $op_2$ \textit{causally depends on} operation $op_1$~\cite{causal-memory, consistency-availability-convergence, cops}, then the ordering must contain $op_1 \leadsto op_2$. Specifically, $op_2$ causally depends on $op_1$ iff.:
\begin{itemize}
    \item $op_1$ and $op_2$ are from the same client and $op_2$ follows $op_1$, or
    \item $op_1$ is a write (or RMW), $op_2$ is a read (or RMW), and $op_2$ returns the written value of $op_1$, or
    \item there is an operation $op'$ s.t. $op_2$ causally depends on $op'$ and $op'$ causally depends on $op_1$ (transitivity).
\end{itemize}

For instance, the following is an SO ordering that satisfies CASL (but not RT), given the same example timeline of \S\ref{sec:relationship-rt}:
\begin{center}
\tikz \graph [grow right sep=0.2cm] {
    b/\opw{e}{x}{3} -> a/\opw{c}{x}{1} -> d/\opr{d}{x}{1} -> e/\opw{d}{y}{3} -> f/\opr{e}{y}{3} -> c/\opw{c}{x}{2};
};
\end{center}

\noindent Notice that \opw{c}{x}{2} ends before \opr{d}{x}{1} starts in physical time, yet \opw{c}{x}{2} $\nleadsto$ \opr{d}{x}{1} in the ordering.

Given this particular final CASL ordering, we can observe that $e$'s read \opr{e}{y}{3} causally depends on $d$'s write \opw{d}{y}{3} (and therefore, transitively, depends on $d$'s read \opr{d}{x}{1} and thus $c$'s write \opw{c}{x}{1}). Meanwhile, it has no interference with $c$'s second write \opw{c}{x}{2}. In other words, in this particular ordering result produced by the service, the potential ``cause'' of $e$ reading value 3 out of $y$ traces back to $c$'s write of value 1 to $x$, but is so far considered irrelevant with $c$'s second write of value 2.

We can in fact visualize the causal dependencies captured by this ordering by drawing arrows that represent potential \textit{causality} between operations on the timeline:
\begin{center}
\tikz \graph [no placement, empty nodes, nodes={circle, fill, scale=0.3}] {
    c1[x=-0.5,y=0,fill=white,scale=3.5,as=$c:$];
    a[x=0,y=0];
    b[x=1.2,y=0];
    abb[x=0.6,y=-0.1,transparent];
    abr[x=1.1,y=0.2,transparent];
    c[x=1.5,y=0];
    d[x=2.7,y=0];
    cdl[x=1.6,y=0.2,transparent];

    c2[x=-0.5,y=-0.8,fill=white,scale=3.5,as=$d:$];
    i[x=3,y=-0.8];
    j[x=3.8,y=-0.8];
    ijl[x=3.1,y=-0.6,transparent];
    ijr[x=3.7,y=-0.6,transparent];
    k[x=4.2,y=-0.8];
    l[x=5.4,y=-0.8];
    kll[x=4.3,y=-0.6,transparent];
    klb[x=4.8,y=-0.9,transparent];

    c3[x=-0.5,y=-1.6,fill=white,scale=3.5,as=$e:$];
    g[x=0.7,y=-1.6];
    h[x=1.9,y=-1.6];
    ghr[x=1.8,y=-1.4,transparent];
    m[x=5.7,y=-1.6];
    n[x=6.5,y=-1.6];
    mnl1[x=5.8,y=-1.3,transparent];
    mnl2[x=5.8,y=-1.4,transparent];
    
    a --["\ptopw{c}{x}{1}"] b;
    c --["\ptopw{c}{x}{2}"] d;

    i --["\ptopr{d}{x}"] j;
    k --["\ptopw{d}{y}{3}"] l;
    
    g --["\ptopw{e}{x}{3}"] h;
    m --["\ptopr{e}{y}"] n;

    abr ->[color=green!70!black,line width=1pt] cdl;
    ijr ->[color=green!70!black,line width=1pt] kll;
    ghr ->[color=green!70!black,line width=1pt] mnl2;
    abb ->[color=green!70!black,line width=1pt,bend right=10] ijl;
    klb ->[color=green!70!black,line width=1pt,bend right=20] mnl1;
};
\end{center}

The following is another valid ordering that is CPO and CASL on the same timeline example; here, \opr{d}{x}{\{1,3\}} observes \opw{e}{x}{3}, setting up an additional causal dependency from \ptopw{e}{x}{3} to \ptopr{d}{x}:
\begin{center}
\tikz \graph [grow right sep=0.2cm] {
    {a/\opw{c}{x}{1}, b/\opw{e}{x}{3}} -> d/\opr{d}{x}{\{1,3\}} -> e/\opw{d}{y}{3} -> f/\opr{e}{y}{3} -> c/\opw{c}{x}{2};
};
\end{center}

CASL is weaker than RT. For each client, its own operations still exhibit the same order as how the client issues them. Across different clients, however, CASL is less restrictive. An operation $op_2$ (or a group of operations) from a client can be reordered before another operation $op_1$ from a different client, even though $op_1$ is ahead of $op_2$ in physical time, as long as $op_2$ has not causally observed $op_1$. This allows certain operations to be processed concurrently without knowing the existence of each other.

\paratitle{Session Guarantees} A popular approach to interpreting causality is to think from each client's perspective (termed a \textit{session}~\cite{session-guarantees}) and decompose the CASL constraint into four \textit{session guarantees}:
\begin{itemize}
    \item \textit{Read My Writes}: if a write $op_1$ and a read $op_2$ are from the same client and $op_2$ follows $op_1$, then $op_2$ must observe $op_1$.
    \item \textit{Monotonic Writes}: writes by a client must happen in the same order as they are issued by the client.
    \item \textit{Monotonic Reads}: if two reads are from the same client, then the latter read cannot observe an older state prior to what the former read has observed. This means if a client issues a read $op_1$ followed by another read $op_2$, then $op_2$ must be ordered after all writes that $op_1$ observes.
    \item \textit{Writes Follow Reads}, i.e., \textit{Session Causality}: if a client issued a read $op'$ that observed a write $op_1$, and later issues a write $op_2$, then $op_2$ must become visible after $op_1$. In this paper, we assume a slightly stricter (but functionally equivalent) version of this guarantee, where $op_2$ must be ordered after the read $op'$ itself\footnote{Having the slightly stricter version of \textit{Writes Follow Reads} allows us to simplify the notion of causality and use a single ordering instead of two (i.e., \textit{visibility order} and \textit{arbitration order}~\cite{non-transactional-summary}) to define all the selected consistency levels on the SOP model.}.
\end{itemize}

The CASL guarantee can be defined exactly as the combination of the above four session guarantees~\cite{session-causality-to-causal-consistency, jepsen-website}.

\subsubsection{First-In-First-Out (FIFO)}

The FIFO guarantee further relaxes CASL by removing write causality dependencies across clients. Specifically, if a read operation $op_r$ from client $c$ observes a write $op_w$ by a different client, now write operations from client $c$ following $op_r$ are allowed to be ordered before $op_r$ and $op_w$. In other words, writes by different clients do not have to maintain their causality order any more.

For instance, the following is an SO ordering that satisfies FIFO (but not CASL), given the same example timeline of \S\ref{sec:relationship-rt}:
\begin{center}
\tikz \graph [grow right sep=0.2cm] {
     b/\opw{e}{x}{3} -> e/\opw{d}{y}{3}-> f/\opr{e}{y}{3} -> a/\opw{c}{x}{1}  -> d/\opr{d}{x}{1} -> c/\opw{c}{x}{2};
};
\end{center}

\noindent Notice that \opw{d}{y}{3} is now ordered before \opw{c}{x}{1} and \opr{d}{x}{1}, breaking the causality chain. Imagine that another client $f$ is reading objects $x$ and $y$; it may then observe $d$'s write to $y$ before seeing $c$'s write to $x$. This may lead to counter-intuitive results for client applications, e.g., showing a user some updated private data before knowing that the user has been removed from the access control list (although the update was made after the ACL removal).

The name FIFO comes from the following analogy: writes from each client are observed by everyone in the same order as they are issued by the client, as if each client pushes its own writes into a separate FIFO queue; meanwhile, writes from different clients are not coordinated with each other by reads.

The FIFO guarantee can be defined exactly as the combination of the \textit{Read My Writes}, \textit{Monotonic Writes}, and \textit{Monotonic Reads} session guarantees~\cite{jepsen-website}. It relaxes CASL by removing \textit{Writes Follow Reads}.

\subsubsection{None Relationship}

An ordering could, of course, place no restrictions on the relative positions of operations. In this case, operations issued by the same client may get arbitrarily reordered. Writes by the same client may be visible to another client in a different order than issued, and a client's read may fail to observe its own preceding write.

This level of relationship constraint demands the least amount of synchronization across operations. Every operation may be processed in a completely asynchronous manner.

\section{Consistency Levels}
\label{sec:consistency-levels}

\begin{figure}[t]
    \centering
    \includegraphics[width=0.98\columnwidth]{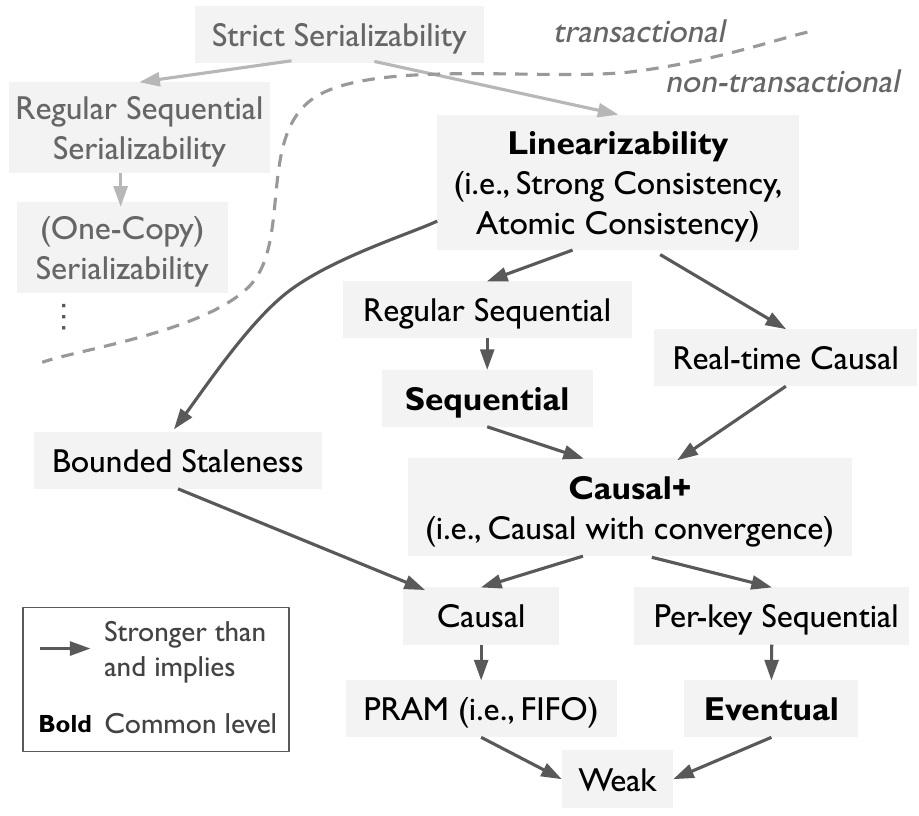}

    \vspace{-5pt}
    \caption{Hierarchy of Selected Consistency Levels.}
    \label{fig:levels-hierarchy}
\end{figure}

\begin{table}[t]
    \centering
    \begin{tabular}{c|c|c}
        \hline
        \textbf{Consistency Level} & \textbf{Convergence} & \textbf{Relationship} \\ \hline \hline
        \textbf{Linearizability} & SO & RT \\ \hline
        Regular Sequential & SO & RT-W \& CASL-R \\ \hline
        \textbf{Sequential} & SO & CASL \\ \hline
        Bounded Staleness & NPO & Bounded-CASL \\ \hline
        Real-time Causal & CPO & RT$'$ \\ \hline
        \textbf{Causal+} & CPO & CASL \\ \hline
        Causal & NPO & CASL \\ \hline
        PRAM & NPO & FIFO \\ \hline
        Per-key Sequential & CPO & CASL-per-key \\ \hline
        \textbf{Eventual} & CPO & None \\ \hline
        Weak & NPO & None \\ \hline
    \end{tabular}

    \vspace{2pt}
    \caption{Ordering Validity Constraints of Consistency Levels.}
    \vspace{-12pt}
    \label{tab:validity-constraints}
\end{table}

We present the hierarchy of useful consistency levels and dissect each level's ordering validity constraints. We first explain the most common consistency levels, namely \textit{linearizability}, \textit{sequential consistency}, \textit{causal+ consistency}, and \textit{eventual consistency}, followed by more subtle levels. We provide examples along the way to help demonstrate their practical differences, and mention representative protocols and systems belonging to each level.

Figure~\ref{fig:levels-hierarchy} presents the hierarchy of selected consistency levels. Arrows represent a ``stronger than'' relationship, where the source level is strictly more restrictive than and thus implies the destination level. Table~\ref{tab:validity-constraints} defines all these consistency levels in a condensed manner by listing their ordering validity constraints.

\subsection{Linearizability}

The strongest non-transactional consistency level is \textit{linearizability}, as defined by Herlihy and Wing in~\cite{linearizability}. In our model, a \textit{linearizable} ordering can be defined as one that satisfies both SO and RT constraints given a physical timeline. It is a serial total order where each operation is ordered before all operations that start after its acknowledgment in real time. A service that provides linearizability is one that always yields a linearizable ordering.

Such a service must maintain some form of a serial log of all operations, where each operation has a specific relative position w.r.t. others. All clients agree on that same order of operations. Furthermore, the service must keep a record of the acknowledgment of each operation, so as to properly order all operations that start after its acknowledgment to satisfy the real-time property.

Linearizability is often referred to as \textit{strong consistency}, due to the fact that it is the strongest possible non-transactional consistency level. Linearizability is sometimes also referred to as \textit{atomic consistency}~\cite{linearizability, parallel-memory-consistency-survey}, because a service that provides linearizability appears to be a piece of shared memory where every client operation is an atomic memory operation. This convenient atomicity semantic makes linearizability one of the easiest consistency levels to reason about and verify against; we can just think of the service as a single piece of atomic memory and apply client operations as they arrive, ignoring all the internal details about complicated distributed system implementation.

\paratitle{State Machine Replication (SMR)} Since the ordering is a serial total order, it is natural to model the object pool as a \textit{state machine} and model client operations as state-transfer \textit{commands}. The service acts as a coordinated set of replicated state machines (typically by replicating the log of operations) and applies committed commands in the decided serial order. This resembles the well-known \textit{State Machine Replication} (SMR) approach~\cite{state-machine-approach, impl-reliable-distributed-multiprocess}, which is widely used in modeling distributed replication systems\footnote{We would like to clarify another closely related term -- \textit{consensus}. A consensus protocol, e.g. Paxos~\cite{paxos-parliament, paxos-made-simple} and others~\cite{ben-or-algorithm, randomized-consensus}, operates at a lower level than a replication protocol; it is used to achieve agreement on a single value (or a sequence of values in optimized variants) among a set of message-passing processes. An SMR protocol, e.g. Multi-Paxos~\cite{paxos-made-simple} or Raft~\cite{raft}, builds atop or inherently integrates a consensus protocol. However, previous literature often extends consensus to include SMR~\cite{raft}.}.

Our \textit{Shared Object Pool} (SOP) model is equivalent to the SMR model if we put some restrictions on both sides. Specifically, an SOP model where only SO orderings are accepted is equivalent to an SMR model where the state is a collection of read-write objects. The SMR model is more expressive than the SOP model in the aspect that it allows more general state machines with custom states and custom commands, not only reads and writes. SOP is more expressive than SMR in the aspect that it inherently allows partial orderings, which helps us incorporate consistency levels that do not guarantee SO.

\paratitle{Protocols \& Systems} Linearizability is the predominant consistency level adopted by critical replication systems built atop SMR protocols. Classic protocols include Chain Replication~\cite{chain-replication}, Multi-Paxos~\cite{paxos-made-simple} and its many variants/optimizations~\cite{cheap-paxos, fast-paxos, vertical-paxos, fpaxos, viewstamped-replication, raft, mencius, epaxos, allconcur, wpaxos, atlas, scalable-smr, insanely-scalable-smr, compartmentalization, consistency-aware-durability, skyros-nil-externality, copilots, rs-paxos, craft, chainpaxos}, Byzantine fault-tolerant protocols~\cite{pbft, hotstuff, qu-protocol, hq-replication}, and others~\cite{rifl, rabia, nopaxos, flair, speculative-paxos, tapir} (some with advanced hardware assumptions). Systems incorporating SMR components include lock/coordination services~\cite{chubby, corfu, tango}, distributed cloud databases~\cite{spanner, cockroachdb, tidb, scylladb, foundationdb, amazon-aurora, farm, rqlite}, and metadata services of large-scale storage systems~\cite{gaios, gfs, etcd, firescroll, ukharon}.

\subsection{Sequential Consistency}
\label{sec:sequential-consistency}

\textit{Sequential consistency}, as originally defined by Lamport in the context of a multiprocessor computer~\cite{sequential-consistency}, means that all clients agree on the same \textit{sequence} of operations applied by the service, where operations from each client appear in the same order as issued by the client. In our model, a service that provides sequential consistency always gives an ordering that is SO and CASL\footnote{Viotti and Vukolić gave a formal formula of sequential consistency that conjuncts SO with PRAM (instead of CASL as in our definition)~\cite{non-transactional-summary}. However, we believe the formula is an erratum and deviates from their text, which reads: ``the realtime ordering of operations invoked by the same process is preserved.'' Their discussion indicates a conjunction with \textit{processor consistency}, which aligns with our CASL constraint.} for any physical timeline workload.

Compared to linearizability, since the ordering does not have to be RT, sequential consistency allows the service to move an operation (or a group of operations) backward in time, reordering it before another group that does not causally precede it. This property is sometimes referred to as \textit{unstable ordering}~\cite{active-quorum-systems, gryff}, in contrast to \textit{stable ordering} provided by linearizability.

For example, given the following physical timeline:
\begin{center}
\tikz \graph [no placement, empty nodes, nodes={circle, fill, scale=0.3}] {
    c1[x=-0.5,y=0,fill=white,scale=3.5,as=$c:$];
    a[x=0,y=0];
    b[x=1.2,y=0];

    c2[x=-0.5,y=-0.8,fill=white,scale=3.5,as=$d:$];
    k[x=2,y=-0.8];
    l[x=3.2,y=-0.8];
    m[x=3.6,y=-0.8];
    n[x=4.6,y=-0.8];
    
    a --["\ptopw{c}{x}{1}"] b;
    
    k --["\ptopw{d}{x}{2}"] l;
    m --["\ptopr{d}{x}"] n;
};
\end{center}

A linearizable ordering must be SO and RT:
\begin{center}
\tikz \graph [grow right sep=0.3cm] {
    a/\opw{c}{x}{1} -> b/\opw{d}{x}{2} -> c/\opr{d}{x}{2};
};
\end{center}

While a sequentially consistent protocol is allowed to give the following ordering that is SO and CASL:
\begin{center}
\tikz \graph [grow right sep=0.3cm] {
    b/\opw{d}{x}{2} -> a/\opw{c}{x}{1} -> c/\opr{d}{x}{1};
};
\end{center}

\noindent The reordering is allowed because client $d$ did not issue any read on object $x$ before \opw{d}{x}{2} that observed value 1 written by client $c$. Therefore, there is no causal dependency from client $c$'s write \opw{c}{x}{1} to client $d$'s write \opw{d}{x}{2}.

At first glance, it may be hard to tell the exact differences between linearizability and sequential consistency. Attiya and Welch presented a quantitative analysis of the performance implications of these two levels, showing that linearizability is strictly more expensive to implement than sequential consistency for common object types in systems without perfectly synchronized clocks~\cite{sequential-vs-linearizability}. But what semantic power do we lose by relaxing the real-time guarantee? The following paragraphs explain three practical implications: 1) sequential consistency does not capture external causality dependencies, 2) sequential consistency is non-local, and 3) it takes extra care to add read-modify-write (RMW) operation support to a sequentially-consistent protocol.

\paratitle{External Causality Dependencies} So far we have assumed that all clients communicate only with the service and there are no \textit{external} communication channels between clients that bypass the service, as depicted in Figure~\ref{fig:sop-model}. However, in real distributed systems such as cloud databases~\cite{cloud-oltp-eval, amazon-aurora, snowflake, cornus}, clients of a replicated storage service may be part of a higher-level system. It is not uncommon for clients to coordinate with each other through external causality dependencies, which are impossible for the service to capture without preserving real-time dependencies.

\begin{figure}[t]
    \centering
    \includegraphics[width=0.86\columnwidth]{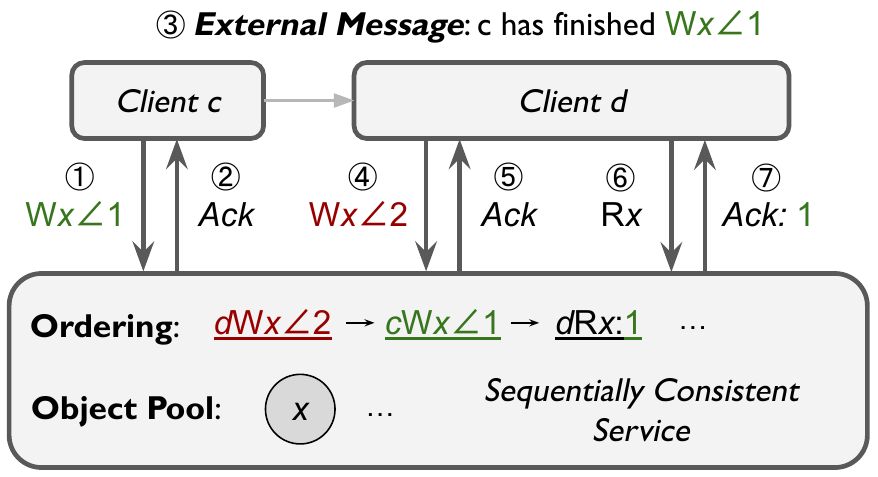}
    
    \vspace{-3pt}
    \caption{Demonstration of External Causality Dependencies.}
    \label{fig:sequential-external-causality}
\end{figure}

In the example depicted by Figure~\ref{fig:sequential-external-causality}, client $c$ first issues a write of value 1 to object $x$ and waits for its acknowledgment. It then sends a message to client $d$ through an external inter-client channel saying ``I have finished my write to $x$ and you can go ahead to operate on $x$.'' Client $d$ then issues its own write of value 2 and expects to read out 2 afterwards. However, since the message from $c$ to $d$ is external to the service, a sequentially consistent service may reorder $d$'s write ahead of $c$'s, and return value 1 for $d$'s read.

A service that provides linearizability will be able to capture such implicit external dependencies because of the real-time property, as \opw{d}{x}{2} starts after \opw{c}{x}{1}'s acknowledgment in physical time\footnote{Note that this is not to be confused with the \textit{external consistency} property in distributed transaction processing systems~\cite{info-storage-decentralized, spanner}, which means that transactions are serialized into the same order as their commit order.}. In contrast, weaker levels that do not honor real time will at best capture \textit{logical} causality, the normal definition of causality (CASL) as described in \S\ref{sec:relationship-casl}.

\paratitle{Implementation Locality} Herlihy and Wing have proven in~\cite{linearizability} that a protocol that implements sequential consistency for each object individually does not necessarily guarantee overall sequential consistency across all operations. Formally, we say that sequential consistency is \textit{non-local}: it is possible for an ordering to be SO and CASL on each object, while not SO or CASL overall.

For example, given the following physical timeline:
\begin{center}
\tikz \graph [no placement, empty nodes, nodes={circle, fill, scale=0.3}] {
    c1[x=-0.5,y=0,fill=white,scale=3.5,as=$c:$];
    a[x=0,y=0];
    b[x=1.2,y=0];
    c[x=1.7,y=0];
    d[x=2.9,y=0];
    e[x=3.4,y=0];
    f[x=4.6,y=0];

    c2[x=-0.5,y=-0.8,fill=white,scale=3.5,as=$d:$];
    g[x=0,y=-0.8];
    h[x=1.2,y=-0.8];
    i[x=1.7,y=-0.8];
    j[x=2.9,y=-0.8];
    k[x=3.4,y=-0.8];
    l[x=4.6,y=-0.8];
    
    a --["\ptopw{c}{x}{1}"] b;
    c --["\ptopw{c}{y}{1}"] d;
    e --["\ptopr{c}{y}"] f;
    
    g --["\ptopw{d}{y}{2}"] h;
    i --["\ptopw{d}{x}{2}"] j;
    k --["\ptopr{d}{x}"] l;
};
\end{center}

The following ordering is SO and CASL on each object (i.e., the \textit{subordering} on object $x$ and $y$ are both SO and CASL), but the overall ordering is CPO and FIFO:
\begin{center}
\tikz \graph [grow right sep=0.3cm] {
    a/\opw{c}{y}{1} -> b/\opw{c}{x}{1} -> c/\opr{c}{y}{2};
    d/\opw{d}{x}{2} -> e/\opw{d}{y}{2} -> f/\opr{d}{x}{1};
    a -> e -> c;
    d -> b -> f;
};
\end{center}

\noindent Notice that given the result of $d$ reading 1 out of $x$ and $c$ reading 2 out of $y$, it is impossible to resolve an SO and CASL ordering across all six operations. This implies that a protocol that guarantees sequential consistency on each object may fail to come up with a global sequence of operations. In fact, such a protocol provides \textit{per-key sequential consistency} (see \S\ref{sec:per-key-sequential-consistency}).

In contrast, a service that provides linearizability on a per-object basis is guaranteed to provide overall linearizability~\cite{linearizability, sequential-vs-linearizability}. We say that linearizability is \textit{local}, allowing modular implementation and verification. The above example can only return value 1 for $c$'s read and value 2 for $d$'s read with such a service.

\paratitle{Support for RMW Operations} A protocol that implements sequential consistency for only read (R) and write (W) operations may take advantage of the unstable ordering of writes to speed up the processing of writes. \textit{Shared register} protocols~\cite{ABD-registers, active-quorum-systems} are the primary examples of this category.

Adding support for read-modify-write (RMW) operations to such protocols is a non-trivial task~\cite{gryff}. In particular, we cannot simply treat RMW operations in the same way as pure writes, because RMWs require a stable base value to determine the result of the read. Systems that demand compare-and-swap (CAS) operations (such as the \textit{LogOnce} operation on shared logs~\cite{cornus}) may have to opt for a service that provides linearizability (or \textit{regular sequential consistency}~\cite{regular-sequential-consistency} as discussed in \S\ref{sec:regular-sequential-consistency}).

\paratitle{Protocols \& Systems} Sequential consistency originates from memory consistency theory~\cite{sequential-consistency, shared-memory-consistency, multiprocessors-simple-consistency-models}. In the context of replicated objects, sequential consistency (or its per-key variant~\cite{pnuts}) is often seen in primary-backup systems~\cite{zookeeper} and message streaming systems~\cite{kafka, redpanda, zeromq} where writes may propagate to readable endpoints after acknowledgment. The transactional form, i.e., \textit{serializability}~\cite{ansi-sql-isolation} plays an indispensable role in database systems.

\subsection{Causal+ Consistency}
\label{sec:causal-plus-consistency}

If a global total order is not required, it may be desirable to further relax sequential consistency and embrace the family of causal consistency levels. Causal consistency stems from the definition of \textit{causal memory}~\cite{causal-memory}. Lloyd et al. pointed out in~\cite{cops} that distributed replication protocols typically implement a slightly stronger version of causal consistency, which they term \textit{causal+ consistency}. It is essentially causal consistency with convergent reads.

In our model, a service that provides causal+ consistency always gives an ordering that is CPO and CASL. Compared to sequential consistency, the ordering does not have to be a serial total order, but instead may leave certain operations from different clients unordered with each other. This opens up opportunities to improve the scalability of a replication protocol. However, all causal dependencies still have to be reflected in the decided ordering.

For example, given the following physical timeline:
\begin{center}
\tikz \graph [no placement, empty nodes, nodes={circle, fill, scale=0.3}] {
    c1[x=-0.5,y=0,fill=white,scale=3.5,as=$c:$];
    a[x=0,y=0];
    b[x=1.2,y=0];
    c[x=1.7,y=0];
    d[x=2.9,y=0];

    c2[x=-0.5,y=-0.8,fill=white,scale=3.5,as=$d:$];
    i[x=1.6,y=-0.8];
    j[x=2.8,y=-0.8];
    k[x=4.5,y=-0.8];
    l[x=5.5,y=-0.8];

    c3[x=-0.5,y=-1.6,fill=white,scale=3.5,as=$e:$];
    m[x=2.4,y=-1.6];
    n[x=3.4,y=-1.6];
    o[x=3.7,y=-1.6];
    p[x=4.9,y=-1.6];
    
    a --["\ptopw{c}{x}{1}"] b;
    c --["\ptopw{c}{y}{1}"] d;
    
    i --["\ptopw{d}{x}{2}"] j;
    k --["\ptopr{d}{y}"] l;

    m --["\ptopr{e}{x}"] n;
    o --["\ptopw{e}{y}{3}"] p;
};
\end{center}

A service that provides causal+ consistency may give the following ordering that is CPO and CASL:
\begin{center}
\tikz \graph [grow right sep=0.3cm] {
    a/\opw{c}{x}{1} -> f/\opw{c}{y}{1};
    b/\opw{d}{x}{2} -> c/\opr{e}{x}{\{1, 2\}} -> d/\opw{e}{y}{3} -> e/\opr{d}{y}{3};
    a -> c;
};
\end{center}

\noindent Notice that \opw{c}{x}{1} and \opw{d}{x}{2} are unordered with each other, and \opr{e}{x}{\{1, 2\}} observes the values of both writes, hence causally depends on both. \opw{e}{y}{3} follows $e$'s read and hence causally depends on both writes as well. \opr{d}{y}{3} observes the result of $e$'s write and hence continues this causal dependency chain, while \opw{c}{y}{1} is dangling and has not been observed by any reader.

\begin{figure}[t]
    \centering
    \includegraphics[width=0.92\columnwidth]{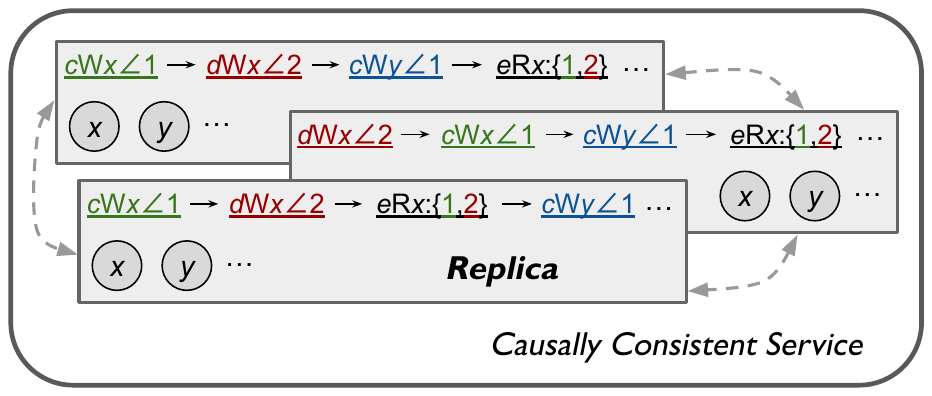}

    \vspace{-3pt}
    \caption{Partial Ordering Interpretation with Replicas.}
    \label{fig:causal-replica-scalability}
\end{figure}

\paratitle{Interpreting A Partial Ordering} Assuming that we are designing a replication protocol atop a set of replica nodes, an intuitive way to interpret a partial ordering in the SOP model is to think from each replica's perspective. Replicas may each maintain a local ordering; different replicas are free to apply different orders for operations that are unordered from the global perspective. Figure~\ref{fig:causal-replica-scalability} demonstrates this perspective.

With a consistency level that always gives an SO ordering, all replicas agree on the same sequence of operations. With a consistency level that allows CPO or NPO ordering, replicas may apply operations in different orders, as long as everyone is coherent with the required validity constraints. This removes the need to coordinate a global sequence for writes that do not causally depend on each other, and is the root source of the scalability and availability benefits of causal+ and weaker consistency levels.

\paratitle{Why Causality} The causal property is desirable in many application scenarios. For example, COPS~\cite{cops} describes a scenario where client $c$ is sharing a photo with client $d$ by first uploading the photo to an image store $s$ and then adding a reference to the photo to the album $a$. Client $d$ then checks $c$'s album and, upon seeing a new reference, goes to fetch the referenced photo:
\begin{center}
\tikz \graph [no placement, empty nodes, nodes={circle, fill, scale=0.3}] {
    c1[x=-0.5,y=0,fill=white,scale=3.5,as=$c:$];
    a[x=0,y=0];
    b[x=1.5,y=0];
    c[x=2,y=0];
    d[x=3.8,y=0];

    c2[x=-0.5,y=-0.8,fill=white,scale=3.5,as=$d:$];
    e[x=4.2,y=-0.8];
    f[x=5.2,y=-0.8];
    g[x=5.5,y=-0.8];
    h[x=6.5,y=-0.8];
    
    a --["\ptopw{c}{s}{\text{photo}}"] b;
    c --["\ptopw{c}{a}{\text{ref}_{\text{photo}}}"] d;
    
    e --["\ptopr{d}{a}"] f;
    g --["\ptopr{d}{s}"] h;
};
\end{center}

For consistency levels that do not honor causal dependencies, such as per-key sequential consistency or eventual consistency, it is possible for $d$ to observe a new reference out of album $a$ but fail to see the new photo from store $s$ (if \opw{c}{s}{\text{photo}} $\nleadsto$ \opr{d}{s}{\text{nil}} in the decided ordering). Causal and thus causal+ consistency prevents this type of counter-intuitive phenomena, because causal dependencies will force \opw{c}{s}{\text{photo}} $\leadsto$ \opr{d}{s}{\text{photo}} since \opw{c}{a}{\text{ref}_{\text{photo}}} $\leadsto$ \opr{d}{a}{\text{ref}_{\text{photo}}}.

\paratitle{Why Convergence} Compared to plain causal consistency, causal+ consistency demands a \textit{convergent conflict resolution} mechanism for conflicting values observed by a read. In other words, all read operations that observe the same set of unordered values on an object must resolve into the same return value. Examples of such conflict resolution mechanisms include \textit{last-writer-wins}, \textit{taking-the-max}, and \textit{taking-the-sum}.

Without the convergence guarantee, causal consistency is allowed to forever return different values for reads on the same object from different clients. This is undesirable in many applications. For example, consider a scenario where two clients, $c$ and $d$, happen to concurrently update the time for a reminder event $t$~\cite{cops}:
\begin{center}
\tikz \graph [no placement, empty nodes, nodes={circle, fill, scale=0.3}] {
    c1[x=-0.5,y=0,fill=white,scale=3.5,as=$c:$];
    a[x=0,y=0];
    b[x=1.5,y=0];
    c[x=2.4,y=0];
    d[x=3.4,y=0];

    c2[x=-0.5,y=-0.8,fill=white,scale=3.5,as=$d:$];
    e[x=0.3,y=-0.8];
    f[x=1.8,y=-0.8];
    g[x=2.4,y=-0.8];
    h[x=3.4,y=-0.8];
    
    a --["\ptopw{c}{t}{\text{7pm}}"] b;
    c --["\ptopr{c}{t}"] d;
    
    e --["\ptopw{d}{t}{\text{8pm}}"] f;
    g --["\ptopr{d}{t}"] h;
};
\end{center}

Original causal consistency may yield the following NPO ordering, letting both $c$ and $d$ falsely believe that their own update is the finalized one, even though  they have indeed observed both writes:
\begin{center}
\tikz \graph [grow right sep=0.3cm] {
    a/\opw{c}{t}{\text{7pm}} -> b/\opr{c}{t}{\text{7pm}};
    c/\opw{d}{t}{\text{8pm}} -> d/\opr{d}{t}{\text{8pm}};
    a -> d;
    c -> b;
};
\end{center}

Causal+ consistency guarantees that $c$ and $d$ agree on the same time value after they have observed both writes. Assuming a last-writer-wins conflict resolution policy, the service may check the acknowledgment timestamp of both writes and determine that the reduced value should be 8pm:
\begin{center}
\tikz \graph [grow right sep=0.3cm] {
    a/\opw{c}{t}{\text{7pm}} -> b/\opr{c}{t}{f(\{\text{7pm}, \text{8pm}\})=\text{8pm}};
    c/\opw{d}{t}{\text{8pm}} -> d/\opr{d}{t}{f(\{\text{7pm}, \text{8pm}\})=\text{8pm}};
    a -> d;
    c -> b;
};
\end{center}

With a service that provides linearizability or sequential consistency, conflicts are avoided altogether by enforcing an SO ordering. However, as previous paragraphs have explained, such protocols inherently have a lower scalability upper bound and a lower availability upper bound.

\paratitle{Protocols \& Systems} Causal dependency originates from causal memory models~\cite{causal-memory, session-guarantees}. It has been adopted by replication systems designed to address availability~\cite{lazy-replication, bolt-on-causal-consistency, unistore, tccstore, practi-replication, bayou} and/or scalability~\cite{cops, chainreaction, occult, antipode, practi-replication, bayou} concerns in large-scale cloud systems, while preserving useful causality semantics.

\subsection{Eventual Consistency}

\textit{Eventual consistency}, as the name suggests, is a consistency level that only requires reads on an object to return a consistent value if no updates are being made to the object. There is no relationship constraint between operations, meaning that any pair of operations are allowed to get reordered, let alone preserving causality, in the final ordering. Eventual consistency is widely adopted in high-demand systems where high performance, scalability, and availability outweigh the need for timely consistency.

\paratitle{Eventual (Strong) Convergence} Although eventual consistency is sometimes used interchangeably with weak consistency, it does impose one extra requirement on the service: the decided ordering must be \textit{(strongly) convergent}~\cite{non-transactional-summary}. Writes eventually become visible to all readers, and reads on the object must all return the same value if they have observed the same writes. This corresponds to the \textit{strong eventual} variant defined in previous literature~\cite{eventual-consistency}; in our model, it is captured by the CPO constraint.

For example, given the following physical timeline:
\begin{center}
\tikz \graph [no placement, empty nodes, nodes={circle, fill, scale=0.3}] {
    c1[x=-0.5,y=0,fill=white,scale=3.5,as=$c:$];
    a[x=0,y=0];
    b[x=1.2,y=0];
    c[x=1.6,y=0];
    d[x=2.8,y=0];
    e[x=3.4,y=0];
    f[x=4.4,y=0];

    c2[x=-0.5,y=-0.8,fill=white,scale=3.5,as=$d:$];
    i[x=1.5,y=-0.8];
    j[x=2.7,y=-0.8];
    
    a --["\ptopw{c}{x}{1}"] b;
    c --["\ptopw{c}{x}{2}"] d;
    e --["\ptopr{c}{x}"] f;
    
    i --["\ptopw{d}{x}{3}"] j;
};
\end{center}

An eventually consistent service is allowed to produce the following CPO ordering:
\begin{center}
\tikz \graph [grow right sep=0.3cm] {
    a/\opw{c}{x}{2} -> b/\opw{c}{x}{1} -> c/\opr{c}{x}{\{1, 3\}};
    d/\opw{d}{x}{3} -> c;
};
\end{center}

\noindent Notice that \opw{c}{x}{2} is allowed to be ordered before \opw{c}{x}{1}, violating the FIFO property. In real implementations, eventually consistent systems typically process every write operation in an asynchronous manner to maximize concurrency. Also notice that \opr{c}{x}{\{1, 3\}} must return a convergent value over the set $\{1, 3\}$.

\paratitle{Quiescent Consistency} A closely related, vaguely defined term is \textit{quiescent consistency}~\cite{art-of-multiprocessor-programming}. In a commonly accepted definition, special periods of physical time called \textit{quiescence period} are identified, during which no write operations are happening. All operations acknowledged ahead of the period are ordered before those that start after the period. Quiescent consistency is weaker than eventual consistency, because it effectively makes no guarantees at all if a system-wide quiescence period never appears~\cite{non-transactional-summary}.

\paratitle{Protocols \& Systems} Eventual consistency is widely adopted by web-scale systems in the form of gossiping protocols and anti-entropy propagation~\cite{grapevine, dynamo, dynamodb, cassandra}. These systems value performance and scalability greatly and can tolerate inconsistencies. A notable line of research related to eventual consistency is on Conflict-free Replicated Data Types (CRDTs)~\cite{crdts, crdt-study, calvin-transactions}.

\subsection{Other Consistency Levels}

In this section, we briefly describe the rest of the selected consistency levels other than the four most common ones. These levels explore different combinations of convergence and (variations of) relationship constraints to refine the consistency level hierarchy.

\subsubsection{Regular Sequential Consistency}
\label{sec:regular-sequential-consistency}

Helt et al. formalized the notion of \textit{regular sequential consistency} in a recent work~\cite{regular-sequential-consistency}. It takes the middle ground between linearizability and sequential consistency. It combines the strengths of both by imposing different levels of relationship constraints for \textit{read-only} operations versus write operations. Specifically, all writes (and RMWs) must honor the real-time property (denoted RT-W), while read operations are allowed to travel back in time as long as they still honor causality (denoted CASL-R).

For example, given the following physical timeline:
\begin{center}
\tikz \graph [no placement, empty nodes, nodes={circle, fill, scale=0.3}] {
    c1[x=-0.5,y=0,fill=white,scale=3.5,as=$c:$];
    a[x=0,y=0];
    b[x=1.2,y=0];
    c[x=3.2,y=0];
    d[x=4.2,y=0];

    c2[x=-0.5,y=-0.8,fill=white,scale=3.5,as=$d:$];
    i[x=1.6,y=-0.8];
    j[x=2.8,y=-0.8];
    
    a --["\ptopw{c}{x}{1}"] b;
    c --["\ptopr{c}{x}"] d;
    
    i --["\ptopw{d}{x}{2}"] j;
};
\end{center}

A service that provides regular sequential consistency may give the following SO ordering, where $c$'s read travels back in time:
\begin{center}
\tikz \graph [grow right sep=0.3cm] {
    a/\opw{c}{x}{1} -> b/\opr{c}{x}{1} -> c/\opw{d}{x}{2};
};
\end{center}

\paratitle{Invariant-equivalence to Linearizability} It is shown that regular sequential consistency is \textit{invariant-equivalent} to linearizability~\cite{regular-sequential-consistency}, meaning that: \circleb{1} it is \textit{local} (see \S\ref{sec:sequential-consistency}) and \circleb{2} it inherently supports RMW operations thanks to stable ordering of writes. However, it does not guarantee to capture \textit{external causality dependencies}, making it still slightly weaker than linearizability. If external causality is not an issue, a linearizable replication system can seamlessly adopt regular sequential consistency to improve the performance of read-only operations.

The transactional version of this consistency level is \textit{regular sequential serializability}~\cite{regular-sequential-consistency}, where read-only transactions are allowed to get reordered in the serialized sequence, while all other transactions must honor RT. Similar properties have been exploited in transactional database systems that use \textit{Timestamp Ordering} (T/O) optimistic concurrency control mechanisms~\cite{tictoc}.

\subsubsection{Real-time Causal Consistency}

\textit{Real-time causal consistency} is a strengthening of causal+ consistency by bringing back a relaxed version of the real-time property. On top of causal+, real-time causal further requires that: if operation $op_1$ is acknowledged before the start of $op_2$ in physical time, then $op_2 \nleadsto op_1$ in the final ordering. Notice that this is a weaker constraint than what we have defined as RT, since RT would enforce $op_1 \leadsto op_2$. We denote this weaker constraint RT$'$.

Assuming that the system is composed of a set of symmetric message-passing replica nodes, Mahajan et al. have proven in~\cite{consistency-availability-convergence} that real-time causal consistency is the strongest possible level that is achievable in an \textit{always-available, one-way convergent} system (which is implied by our definition of \textit{sticky available} in \S\ref{sec:availability-guarantees}).

\paratitle{Fork-based Consistency Models} A family of fork-based consistency models has been developed to deal with Byzantine faults in a system containing untrusted components. For example, a \textit{fork-linearizable} system ensures that if any two replicas have observed different orderings (i.e., \textit{forked} by an adversary), then their writes will never be visible to each other afterwards (i.e., they cannot be \textit{joined} again). \textit{Fork causal consistency} is a family of consistency levels that weaken causal consistency to tolerate Byzantine replicas and enforce causal consistency among correct replicas~\cite{depot}.

\subsubsection{Causal Consistency}
\label{sec:causal-consistency}

Causal and causal+ consistency have been explained in \S\ref{sec:causal-plus-consistency}. As a recap, a service that provides \textit{causal consistency} must give an ordering that is NPO and CASL given any physical timeline workload. Such an ordering captures all the potential causality dependencies between operations, but does not demand convergent conflict resolution, meaning that different clients are allowed to forever retrieve different values from reads on the same object.

As mentioned in \S\ref{sec:relationship-casl}, causal consistency can be defined exactly as the combination of the four session guarantees~\cite{session-causality-to-causal-consistency, jepsen-website}.

\subsubsection{Bounded Staleness}
\label{sec:bounded-staleness}

Although causal consistency enables the powerful abstraction of causal dependency, it does not provide any guarantee on the ``timeliness'' of when writes become visible to reads. \textit{Bounded staleness} is a vaguely-defined family of consistency levels that typically strengthen causal consistency by adding \textit{recency} guarantees~\cite{cosmosdb-consistency-levels}.

Bounded staleness levels put an extra constraint on the \textit{delay} between the acknowledgment of a write by client $c$ on object $x$ and when reads from other clients on $x$ must reflect the effect of the write. The delay constraint may be expressed in the following ways: 1) at most $j$ more write operations by client $c$, or 2) at most $k$ more updates on object $x$, or 3) at most a physical time interval $t$, or 4) a mixture of the three, e.g., whichever is reached first. We use the name Bounded-CASL to broadly refer to the combination of the CASL relationship guarantee with any delay constraint.

Because of the extra delay constraint, bounded staleness levels are incomparable with both sequential and causal levels, because they both do not express any recency requirements.

\subsubsection{PRAM Consistency}
\label{sec:pram-consistency}

\textit{Pipeline Random Access Memory} (PRAM) consistency~\cite{pram-consistency}, or simply \textit{FIFO consistency}, is a weaker consistency level than causal consistency, where causality across clients is not captured. It was originally defined for shared memory systems. In our framework, it is a consistency level that requires NPO and FIFO ordering.

Using the notion of session guarantees, PRAM consistency can be defined exactly as the combination of \textit{Monotonic Writes}, \textit{Monotonic Reads}, and \textit{Read My Writes}~\cite{jepsen-website}. It does not enforce \textit{Writes Follow Reads}, hence not capturing cross-client causality.

\paratitle{Consistent Prefix} The combination of \textit{Monotonic Writes} and \textit{Monotonic Reads} are sometimes referred to as \textit{Consistent Prefix}~\cite{cosmosdb-consistency-levels}. This name comes from the fact that, for every writer, all clients will observe a monotonically-growing prefix of its writes.

Although Figure~\ref{fig:levels-hierarchy} does not include consistent prefix because of its vague definition, we can derive a strength rank of this level w.r.t. bounded staleness, causal, and PRAM consistency: any Bounded Staleness configuration $>$ Causal $>$ PRAM $>$ Consistent Prefix.

\subsubsection{Per-key Sequential Consistency}
\label{sec:per-key-sequential-consistency}

As \S\ref{sec:sequential-consistency} pointed out, sequential consistency is \textit{non-local}, meaning that a protocol that enforces SO and CASL ordering on a per-object basis (termed CASL-per-key) does not necessarily guarantee a global SO and CASL ordering across all operations. In fact, such a protocol implements \textit{per-key sequential consistency}.

This consistency level was first studied in the PNUTS system~\cite{pnuts}, a highly-concurrent data serving system that provides per-record consistency. However, modern distributed systems typically have complicated client-side logic layered on top of a non-transactional object store, where each client is interested in more than one object. This makes the object-key-oriented consistency level less appealing than session-oriented causality levels. The photo-album case described in \S\ref{sec:causal-plus-consistency} would be a good example that demonstrates the limitations of per-key sequential consistency.

\subsubsection{Weak Consistency}

\textit{Weak consistency}\footnote{Weak consistency is irrelevant to \textit{weak ordering} in shared memory systems~\cite{memory-consistency-models, multiprocessors-simple-consistency-models}.} is at the bottom of the consistency level spectrum and is weaker than all other consistency levels. In our model, weak consistency can be defined as enforcing an NPO and None-relationship ordering. It can simply be interpreted as ``providing no consistency guarantees at all''.

\subsubsection{Mixed/Hierarchical Consistency Levels}

So far, we have assumed a single conceptual storage service without making any assumptions on the internal implementation of the service. Real distributed systems may, however, contain multiple layers or scopes of sub-services, each providing a different consistency level semantic. For example, CosmosDB~\cite{cosmosdb-consistency-levels} provides a stronger consistency guarantee for clients within the same \textit{region} than those distributed across multiple regions, effectively exposing a 2-layer consistency model. Given the implementation details of a system, we can always define mixed or hierarchical consistency levels composed of multiple basic levels.

Yu and Vahdat~\cite{tact-continuous, continuous-consistency} proposed a continuous consistency model for replicated services, where consistency is defined as a 3-tuple, (\textit{numerical error}, \textit{order error}, and \textit{staleness}), named a \textit{conit}. This leads to a fairly fine-grained consistency spectrum and allows applications to dynamically balance consistency and performance.

\subsubsection{Memory Consistency Models}

Distributed replication consistency is tightly related to early works in multiprocessor shared memory consistency. Hill defined \textit{hardware memory consistency model} as the interface contract for shared memory, where instructions may be executed out-of-order~\cite{multiprocessors-simple-consistency-models}. Memory consistency models and techniques such as \textit{weak ordering}, \textit{acquire/release consistency}, \textit{entry consistency}, \textit{cache coherence}, and \textit{memory fences/barriers}~\cite{memory-consistency-models, multiprocessors-simple-consistency-models} are out of the scope of this paper.

\section{Availability Guarantees}
\label{sec:availability-guarantees}

Besides consistency, \textit{availability} is also an important part of the interface contract between a distributed storage service and clients. Availability is not implementation-oblivious; the meaning of fault-tolerance and availability can only be defined given a specific system model. In this section, we consider a simple system of symmetric replicas and analyze the best possible availability guarantee that each consistency level can provide in such a system.

\subsection{Symmetric Replicas System Model}

We consider a fault-tolerant system implementation of the object store service composed of a set of \textit{symmetric replica servers}, similar to what Figure~\ref{fig:causal-replica-scalability} depicts. Each replica node holds a complete copy of all objects and can communicate with any other replica through messages over the network. Clients establish connections to one (or more) replica(s), issue operations, and wait for acknowledgments.

\paratitle{Data Partitioning} Since we only consider non-transactional workloads, this symmetric model can be easily extended to incorporate \textit{data partitioning} (or called \textit{partial replication}), where each node is responsible for a subset of objects. For each object, only the set of nodes that hold the object is under consideration for availability.

\paratitle{Client-side Caching} A client may act as a partial replica server by doing \textit{client-side coherent caching} w.r.t. the consistency level for its reads and writes~\cite{bolt-on-causal-consistency, session-guarantees}. In this case, we count the client itself as a valid partial replica.

\subsection{Meaning of Availability}

Consider a non-Byzantine fail-stop setting with an asynchronous network~\cite{paxos-made-simple}. We say a system of symmetric replicas provides \textbf{availability} if, in the presence of arbitrarily long network partitions between arbitrary replicas, every client that can connect to one (or a specific set of) non-failing replica(s) of an object can get valid acknowledgments for all operations it issues on that object.



\paratitle{Availability Levels} We consider three coarsely-defined levels~\cite{jepsen-website}:
\begin{itemize}
    \item \textit{Totally available}: every client that can contact \textit{at least one} non-failing replica of an object eventually receives responses that honor the consistency level for operations on that object.
    \item \textit{Sticky available}: a client maintains \textit{stickiness} if it keeps contacting the same replica for all of its operations on an object. The system is sticky available if every client that sticks to a non-failing replica of an object eventually receives responses that honor the consistency level for operations on that object.
    \item \textit{Weakly available}: the system does not guarantee progress under arbitrary network partitions.
\end{itemize}

Note that the ``weakly available'' category can be further decomposed into finer-grained, protocol-specific availability levels if we can bound the number of failures to a certain quantity. For example, most state machine replication protocols are available when at least a majority of nodes are healthy and connected. Also, extra care needs to be taken to define reasonable transactional availability guarantees~\cite{highly-available-transactions}, which is out of the scope of this paper.

\subsection{Availability Upper Bounds}

The \textit{CAP theorem} states that a distributed system cannot achieve Consistency, Availability, and network Partition-tolerance all at the same time~\cite{towards-robust-distributed}. This informal description is often taken in an overly restrictive form. A more precise statement would be that a distributed system cannot achieve linearizability, total/sticky availability, and tolerance to full network partitioning all at the same time. This statement has been proven by Gilbert and Lynch~\cite{cap-theorem}.

By relaxing linearizability to weaker consistency levels, it is often (but not always) possible to derive a replication protocol that guarantees sticky or even total availability under arbitrary network partitions. Table~\ref{tab:availability-upper-bound} lists the availability upper bound of each of the selected consistency levels.

\begin{table}[t]
    \centering
    \begin{tabular}{c|c}
        \hline
        \textbf{Consistency Level} & \textbf{Availability Upper Bound} \\ \hline \hline
        \textbf{Linearizability} & \multirow{4}{*}{Weakly available} \\ \cline{1-1}
        Regular Sequential & \\ \cline{1-1}
        \textbf{Sequential} & \\ \cline{1-1}
        Bounded Staleness & \\ \hline
        Real-time Causal & \multirow{7}{*}{Sticky available} \\ \cline{1-1}
        \textbf{Causal+} & \\ \cline{1-1}
        Causal & \\ \cline{1-1}
        PRAM & \\ \cline{1-1}
        Per-key Sequential & \\ \cline{1-1}
        \multicolumn{1}{l|}{\textit{Session Guarantees}:} & \\
        \multicolumn{1}{l|}{\quad Read My Writes} & \\ \cline{2-2}
        \multicolumn{1}{l|}{\quad Writes Follow Reads} & \multirow{5}{*}{Totally available} \\
        \multicolumn{1}{l|}{\quad Monotonic Reads} & \\
        \multicolumn{1}{l|}{\quad Monotonic Writes} & \\ \cline{1-1}
        \textbf{Eventual} & \\ \cline{1-1}
        Weak & \\ \hline
    \end{tabular}

    \vspace{1pt}
    \caption{Availability Upper Bounds of Consistency Levels.}
    \vspace{-15pt}
    \label{tab:availability-upper-bound}
\end{table}

Most of these availability bounds have been proven in previous literature~\cite{consistency-availability-convergence, highly-available-transactions}. Linearizability, regular sequential consistency, and bounded staleness are obviously weakly available because of the RT constraint or the delay constraint: clients connecting to servers separated on opposite sides of a network partition have no way of knowing the acknowledgment time of operations made on the other side, unless operations on that side are blocked indefinitely. Sequential consistency cannot be sticky available because of its non-locality, as counter-examples similar to the one presented in \S\ref{sec:sequential-consistency} can be constructed; in contrast, per-key sequential is sticky available. Bailis et al. have proven that the writes follow reads, monotonic reads, and monotonic writes session guarantees are totally available, while read my writes requires stickiness~\cite{highly-available-transactions}. Causal and PRAM consistency are therefore both sticky available. Mahajan et al. have proven that real-time causal is as available as causal consistency (given one-way convergence, which is assumed in our model)~\cite{consistency-availability-convergence}. Causal+ is also sticky available following this result. Eventual and weak consistency are both totally available: clients can make progress on any live server.

\paratitle{Limitations} The availability upper bounds presented here are rather coarse-grained and do not capture everything about availability. First, they say nothing about \textit{recency} guarantees, i.e., how stale are read results allowed to be. For example, although causal consistency is sticky available, a network partition may indefinitely prevent writes made on one side from being visible to readers on the other side. Bounded staleness levels would thus all be weakly available in our definition. Second, these availability bounds do not consider \textit{partial} network partitions, where certain pairs of nodes cannot directly communicate with each other, but some indirect multi-hop paths are still available. Alfatafta et al. discussed partial network partitions and mechanisms to exploit indirect paths~\cite{partial-network-partitions}.

\section{Checker Implementation}
\label{sec:checker-impl}

To further demonstrate the uniformity, practicality, and understandability of the SOP model, we apply it to consistency checking. Assume a known number of clients using a key-value store service as depicted throughout the paper. Given a history of client operations as input, our consistency model should be able to decide which consistency levels the service conforms to (according to the specific history) and which levels it certainly already violates.

We use Jepsen, a distributed system testing, fault injection, and analysis toolchain~\cite{jepsen-github} (open-sourced by the same-named company~\cite{jepsen-website}), as the basis of our implementation. Jepsen offers an automated workflow for running real distributed systems, generating client workloads, injecting failures, recording the execution history, and performing consistency and availability analysis.

We implement a consistency levels conformity checker prototype in \textasciitilde1k lines of Rust, using SOP orderings as the underlying mechanism. We add \textasciitilde1k lines of Clojure wrappers to integrate the checker with Jepsen and make it a selectable alternative to the original Knossos linearizability analyzer for key-value operations. The demo can be found at \url{https://github.com/josehu07/jepsen.demo}.

\subsection{Checker Logic}

The checker takes as input from the Jepsen execution stage a \textit{history}, which is a sequence of events where each event is either the invocation or the completion of a client operation. There are three types of operations: read (R), write (W), and compare-and-swap (CAS). The three types correspond to the model's definition described in \S\ref{sec:sop-model}, with CAS being a concrete, representative type of an RMW operation that conditionally writes a new value if passing an equality check on an expected old value.

The checker outputs four flags to indicate whether the given history conforms to the four common consistency levels: linearizable (Linr.), sequential (Seql.), causal+ (Casl+), and eventual (Evtl.). Conveniently, due to the chain ranking across the four levels, satisfying a higher level guarantees all weaker levels. Note that the result of a run is specific to the particular history produced in the run, and the system in general could be at a weaker level than tested.

The internal logic of the checker goes as follows. \circleb{1} It parses the history into a timeline resembling \S\ref{sec:physical-timeline}, stored as a collection of per-client queues of \textit{spans}, where each span represents a specific operation with start and end timestamps. \circleb{2} It repeatedly drains the queues in bulk of concurrent operations, and tries to iterate through all possible constructions of ordering graphs. If a graph satisfying both the convergence and relationship validity constraints of a level is found when all queues in the timeline have been drained, that level is satisfied. The iterative process starts from "easier" graphs (e.g., SO graphs with RT relationship constraint), seeking stronger levels first to terminate early, before moving on to "harder" graphs (e.g., CPO graphs with more flexible relationship constraints). \circleb{3} If all the CPO graph possibilities are exhausted for all chunks of spans, or if any read returns a corrupted or never-seen value, the checker terminates with all flags set to false (meaning weak consistency).

\subsection{Analysis Results}

We run the Jepsen workflow on three representative systems: the etcd key-value store~\cite{etcd}, the ZooKeeper coordination service~\cite{zookeeper}, and the RabbitMQ message broker~\cite{rabbitmq}, with various configurations when relevant. All systems are run with 5 replicas distributed across 5 CloudLab machines~\cite{cloudlab}, and are directed to expose a replicated key-value store service API to clients.

In each run, 10 clients are distributed across the same machines evenly, and generate 30 seconds of workloads with random keys and values at a global 200 ops/sec rate. Network partitioning faults are injected every 10 seconds and last 5 seconds each time. This gives a diverse coverage of the consistency hierarchy across the four most common levels. We do not measure availability.

Table~\ref{tab:checker-results} presents the results of the runs. Our SOP-based checker outputs fine-grained consistency validation results that span the four levels. Jepsen's original Knossos analyzer outputs a binary decision on linearizability only. The results of all six system configurations match what we expected from the system deployments; we explain the deployment modes below.

\paratitle{etcd Modes}
etcd is a Raft-consensus-backed, strongly-consistent key-value store for critical data with transaction support. Default deployment uses quorum reads, following the Raft protocol strictly, and is therefore linearizable. If reads are allowed to be acknowledged before reaching a majority quorum (Stale read), they could miss the latest committed writes, and the service degrades to sequential consistency. We also test a mode where CAS operations are implemented manually as serializable transactions instead of single-point operations, which also brings the overall consistency down to sequential.

\paratitle{ZooKeeper (ZK) Modes}
ZooKeeper is a sequentially-consistent coordination service backed by the ZAB primary-backup protocol. We use ZK through the Avout library, which provides a distributed Clojure \textit{atom} abstraction using ZK as access locks. Although sequentially consistent, triggering a non-linearizable read result is rare (as it requires stale locks to be held in close succession); thus, our test run yielded a linearizable history ($^*$). We also include a mode where a subset of atoms is replaced with local atom references without cross-node communication. This caps those atoms at causal+ consistency.

\paratitle{RabbitMQ (RMQ) Modes}
RabbitMQ is a message queueing and brokerage system. We build a peer-to-peer broadcasting layer using RabbitMQ queues co-located with each node as the communication media between them. This resembles a weakly-consistent key-value store service to clients, where updates are propagated lazily to peers through background announcements.

\paratitle{Performance Limitation}
The main purpose of the checker implementation is to demonstrate the uniformity, expressiveness, and practical relevance of the SOP model. A major limitation lies in brute-force ordering graph construction, which has sub-factorial complexity w.r.t. the number of concurrent writes (bounded by the number of clients), leading to long analysis time and high memory consumption. Checking for linearizability alone takes \textasciitilde1s, on par with Jepsen's in our small-scale tests, but weaker levels require hours to explore. In practice, level-specific algorithms are used~\cite{p-compositionality, jepsen-knossos, porcupine, serializability-np-complete, causal-np-complete, weak-consistency-checking} with auxiliary information from the tested system (such as versions and the believed serialization order).

\begin{table}[t]
    \centering
    \setlength\tabcolsep{4.0pt}
    \begin{tabular}{c|c|c|c|c|c|c}
        \hline
        \multicolumn{2}{c|}{\textbf{System Configuration}} & \multicolumn{4}{c|}{\textbf{SOP-based Checker}} & \textbf{Jepsen} \\
        \hline
        System & Mode & Linr. & Seql. & Casl+ & Evtl. & Knossos \\
        \hline \hline
        \multirow{3}{*}{\textbf{etcd}} & Quorum read & \CIRCLE & \CIRCLE & \CIRCLE & \CIRCLE & Pass \\
        \cline{2-7}
          & Stale read & \Circle & \CIRCLE & \CIRCLE & \CIRCLE & No \\
        \cline{2-7}
          & CAS as txns & \Circle & \CIRCLE & \CIRCLE & \CIRCLE & No \\
        \hline
        \multirow{2}{*}{\textbf{ZK}} & Locked atoms & \CIRCLE$^*$ & \CIRCLE & \CIRCLE & \CIRCLE & Pass \\
        \cline{2-7}
          & Local refs & \Circle & \Circle & \CIRCLE & \CIRCLE & No \\
        \hline
        \textbf{RMQ} & P2P announce & \Circle & \Circle & \Circle & \CIRCLE & No \\
        \hline
    \end{tabular}

    \vspace{1pt}
    \caption{Jepsen Workflow Checker Analysis Results.}
    \vspace{-15pt}
    \label{tab:checker-results}
\end{table}

\section{Conclusion}
\label{sec:conclusion}

This paper presents a unified, practical, and understandable model of non-transactional \textit{consistency} levels in the context of distributed replication systems. We develop an intuitive model -- the shared object pool (SOP) -- and show that concise definitions of useful consistency levels can be constructed out of two types of ordering validity constraints: \textit{convergence} and \textit{relationship}.
We demonstrate the model's expressiveness and practical relevance with a Jepsen-integrated checker implementation. As replicated systems become the cloud-era norm, we believe this paper provides useful guidance to protocol designers and distributed system engineers.

\bibliographystyle{ACM-Reference-Format}
\bibliography{references}

\end{document}